\documentclass[
    twocolumn,
    showpacs,
    nofootinbib,
    preprintnumbers,
    amsmath,
    amssymb,
    showkeys,
    superscriptaddress]{revtex4-1}
\usepackage{epsfig}
\usepackage{dsfont}
\usepackage{color}% Include figure files
\usepackage{graphicx}% Include figure files
\usepackage{dcolumn}% Align table columns on decimal point
\usepackage{bm}% bold math
\usepackage{physics}

\begin{document}
\allowdisplaybreaks

\title{The $1/N$ expansion for stochastic fields in de Sitter spacetime}% Force line breaks with \\

\author{G. Moreau} %\ead{moreau@apc.univ-paris7.fr}
\author{J. Serreau}%\ead{serreau@apc.univ-paris7.fr}%
\affiliation{Universit\'e de Paris, CNRS, Astroparticule et Cosmologie, F-75013 Paris, France.}
\author{Camille No\^us}
\affiliation{Cogitamus Laboratory}

\date{\today}% It is always \today, today,

\begin{abstract}

We propose a $1/N$ expansion of Starobinsky and Yokoyama's effective stochastic approach for light quantum fields on superhorizon scales in de Sitter spacetime.
We explicitly compute the spectrum and the eigenfunctions of the Fokker-Planck operator for a O($N$)-symmetric theory with quartic selfinteraction at leading and next-to-leading orders in this expansion. We obtain simple analytical expressions valid in various nonperturbative regimes in terms of the interaction coupling constant.

\end{abstract}

%\pacs{?????????????}% PACS, the Physics and Astronomy
%\keywords{}%Use showkeys class option if keyword
\maketitle

%\tableofcontents
\section{Introduction}
\label{sec:intro}

The stochastic formalism is a powerful way to access the infrared physics of light quantum fields in slow-roll inflationary backgrounds \cite{Starobinsky:1994bd}. It provides an effective description of the dynamics on superhorizon scales in terms of (coupled) Langevin equations. Correlators can be extracted from this formulation \cite{Starobinsky:1994bd,Markkanen:2019kpv,Moreau:2019jpn,Bounakis:2020jdx}, and have been shown to correctly capture the infrared behavior of the full quantum field theory at leading-infrared-logarithm accuracy \cite{Tsamis:2005hd,Garbrecht:2014dca}. The stochastic formalism coexists with several alternative nonperturbative methods \cite{vanderMeulen:2007ah,Burgess:2009bs,Rajaraman:2010xd,Serreau:2011fu,Akhmedov:2011pj,Beneke:2012kn,Boyanovsky:2012qs,Parentani:2012tx,Kaya:2013bga,Serreau:2013eoa,Gautier:2013aoa,Youssef:2013by,Gautier:2015pca,Boyanovsky:2015tba,Guilleux:2015pma,Moss:2016uix,Prokopec:2017vxx,LopezNacir:2018xto}. 

Focusing on the case of spectator fields with a standard kinetic term\footnote{For discussion of nonstandard kinetic terms in nonlinear $\sigma$ models, see Ref.~\cite{Kitamoto:2018dek}.} in pure de Sitter spacetime, the relevant stochastic dynamics possesses a late time equilibrium state described by a stationary probability distribution whose form is known for an arbitrary potential. This allows one to compute a variety of one point correlators  often with analytic control. Higher order correlators exhibit nontrivial spacetime dependences which can be conveniently expressed in terms of a spectral decomposition involving the eigenvalues and eigenfunctions of the associated Fokker-Planck operator. If those can be computed numerically \cite{Starobinsky:1994bd,Markkanen:2019kpv,Markkanen:2020bfc,Adshead:2020ijf}, it is often of interest to also have some analytic control, for instance, as checks of numerical results, or for comparison with direct QFT calculations \cite{Gautier:2013aoa,Garbrecht:2014dca,Gautier:2015pca,LopezNacir:2018xto,Bounakis:2020jdx}. 

To date, only few explicit analytical results are known concerning the spectrum of the Fokker-Planck operator, even in the case of a simple quartic potential. Of course, when the relevant coupling constant (see below) is small, a systematic perturbative treatment of the eigenvalue problem is feasible. This has been implemented at the first nontrivial orders  both in the case of a positive \cite{Markkanen:2019kpv} and of a negative \cite{Markkanen:2020bfc} square mass for a single scalar field theory. Such perturbative results, however, are not valid in the (phenomenologically relevant) cases of essentially massless fields. Nonperturbative expressions of the three lowest eigenvalues have been obtained from the calculation of various correlators in a $1/N$ expansion for a O($N$)-symmetric theory \cite{Moreau:2019jpn}. 

In the present work, we setup a proper $1/N$ expansion directly at the level of the Fokker-Planck eigenvalue equation for systems with O($N$) symmetry.  In the case of a quartic potential, we obtain simple analytical expressions of all eigenvalues and eigenfunctions both at leading and next-to-leading orders, which reproduce and encompass the results mentioned above. These provide benchmark results, valid for arbitrary value of the coupling (within the validity of the stochastic approach, {\it i.e.}, for light fields), for various quantities of physical interest, such as correlation lengths and times, relaxation and decoherence timescales, or various spectral indices, relevant for phenomenological applications \cite{Hardwick:2017fjo,Martin:2018lin,Markkanen:2019kpv,Adshead:2020ijf}.

In Sec.~\ref{sec:formalism}, we briefly review the stochastic approach for the O($N$) theory and its formulation in terms of an eigenvalue problem for the associated Fokker-Planck operator. We setup the $1/N$ expansion of the problem and present the solution to the eigenvalue problem at leading and next-to-leading orders in Sec.~\ref{sec:LOcomputation}. We discuss our findings together with their physical interpretation and the comparison with previously existing results  in Sec.~\ref{sec:discussion}. Sec.~\ref{sec:concl} summarizes our conclusions. The details of the next-to-leading-order calculation are given in the Appendix \ref{sec:NLOcomputation} and we present some comparison with numerical results in Appendix \ref{sec:appnum}.

\section{Stochastic formalism and Fokker Planck equation}
\label{sec:formalism}

We consider a $N$-component scalar field $\hat\varphi_a$ embedded in the expanding Poincar\'e patch of de Sitter spacetime in $D=d+1$ dimensions, with metric $\dd s^2 =g_{\mu\nu}\dd x^\mu \dd x^\nu =-\dd t^2 + e^{2Ht} \dd \vec x^2$, in terms of the cosmological time $t$ and the comoving spatial coordinates $\vec x$. We set the Hubble scale $H=1$ in the following. With standard notations, the microscopic action reads 
\begin{equation}
    S = -\int \dd[D]x \sqrt{-g} \left\{ \frac{1}{2}\partial_\mu \hat\varphi_a \partial^\mu \hat\varphi_a + \hat V(\hat\varphi_a) \right\}.
    \label{}
\end{equation}
For light fields in units of $H$,\footnote{More precisely, this holds in the regime where the curvature of the potential in field space is small in units of the spacetime curvature.} the quantum fluctuations on superhorizon scales can be described as those of an effective stochastic variable driven by the subhorizon degrees of freedom. On such scales, spatial gradient are negligible and one can treat the problem as a collection of independent Hubble patches described by an appropriate Langevin equation \cite{Starobinsky:1994bd}. Absorbing unimportant numerical factors through the redefinitions $\varphi_a = \hat\varphi_a\sqrt{d\Omega_{D+1}/2}$ and $V = \hat V\Omega_{D+1}/2$, with $\Omega_{n}  = 2\pi^{n/2}/\Gamma(n/2)$, the latter reads
\begin{equation}
    \partial_t\varphi_a + \partial_aV = \xi_a,
    \label{eq:langevin}
\end{equation}
where $\partial_a=\partial/\partial\varphi_a$. Here, the field $\varphi_a$ denotes a spatially averaged field over a Hubble patch and the noise $\xi_a$ reflects the effect of the subhorizon (quantum) fluctuations, which constantly feed this coarse-grained degree of freedom as a result of the gravitational redshift. We refer the reader to the literature \cite{Starobinsky:1994bd,Tsamis:2005hd} for details on the derivation of Eq.~\eqref{eq:langevin}.  Treating the subhorizon sector in the linear approximation and assuming the Bunch-Davies vacuum yields a white Gaussian noise, entirely characterized by the two-point correlator 
\begin{equation}
    \ev{\xi_a(t)\xi_b(t')} = \delta_{ab} \delta(t-t').
    \label{}
\end{equation}
Following standard procedures \cite{Risken:1996}, this stochastic dynamics is equivalently formulated in terms of the following Fokker-Planck equation for the field probability distribution function (PDF) $P\equiv P(\varphi_a,t)$ 
\begin{equation}
    \partial_tP = \partial_a\qty[(\partial_aV) P + \frac12 \partial_aP],
    \label{eq:FPP}
\end{equation}
which can, itself, be reduced to an eigenvalue problem, as we now recall for the case of a potential with O($N$) symmetry. 

First, introduce the reduced PDF $p(\varphi_a,t)$, defined as $P(\varphi_a,t) = e^{-V(\varphi_a)} p(\varphi_a,t)$, in terms of which Eq.~\eqref{eq:FPP} takes the form of the Schr\"odinger-like equation
\begin{equation}
    \partial_tp = \frac12 \Delta_\varphi p - W p,
    \label{eq:FPp}
\end{equation}
where $\Delta_{\varphi}=\partial_{a}\partial_{a}$ and
\begin{equation}
    W = \frac12 \qty[ \qty(\partial_a{V})^2-\Delta_\varphi V  ].
    \label{eq_pot}
\end{equation}
For a O($N$)-symmetric potential, it is convenient to use spherical coordinates in field space and to decompose the angular dependence onto generalized spherical harmonics $Y_{\ell_i}(\theta_i)$, where $\theta_{i=1,\ldots,N-1}$ denote the $N-1$ angular variables in field space and the $\ell_i$ are integers such that $|\ell_1|\le\ell_2\le\ldots\le\ell_{N-1}$. These harmonics diagonalise the angular part of the operator $\Delta_{\varphi}=\partial_\rho^2+(N-1)\partial_\rho/\rho+ \Delta_{S^{N-1}}/\rho^2$ as
\begin{equation}
    \Delta_{S^{N-1}} Y_{\ell_i}(\theta_i) = - \ell(\ell+N-2) Y_{\ell_i}(\theta_i),
    \label{}
\end{equation}
where we have noted $\ell=\ell_{N-1}$ and $\rho=\sqrt{\varphi^2}$. For the purpose of the $1/N$ expansion below, it is convenient to introduce the scaled radial variable and potentials $x = \rho/\sqrt{N}$, $v = V/N$, and $w=W/ N$. We have 
\begin{equation}
    w = \frac1{2N}\qty[N (v')^2-v'' - (N-1) \frac{v'}{x} ],
    \label{eq_potscaled}
\end{equation}
where the prime denotes a derivative with respect to $x$. Seeking solutions to Eq.~\eqref{eq:FPp} of the form $p(\varphi_a,t) = \mathcal{R}(x) Y_{\ell_i}(\theta_i) e^{-\Lambda t}$ yields the eigenvalue problem
\begin{equation}
    -\frac{\mathcal{R}''}{2N} - \frac{N-1}{2Nx} \mathcal{R}' + \qty[\frac{\ell(\ell+N-2)}{2Nx^2} + N w] \mathcal{R} = \Lambda\mathcal{R}.
    \label{eq:radial}
\end{equation}

\section{The $1/N$ expansion}
\label{sec:LOcomputation}

\subsection{Gaussian guidance}

To set up an appropriate $1/N$ expansion we first need to properly control the limit $N\to\infty$ of the theory. This requires one to understand how the various quantities in Eq.~\eqref{eq:radial} scale with $N$ at large $N$. To this aim, it is instructive to consider the exactly solvable case of a purely quadratic potential $v(x) = m^2 x^2/2$, or, equivalently, 
\begin{equation}
    w(x) = - \frac{m^2}2 + \frac{m^4}2 x^2.
    \label{eq:gaussian}
\end{equation}
In that case, Eq.~\eqref{eq:radial} is nothing but the radial Schr\"odinger equation for a symmetric $N$-dimensional harmonic oscillator with unit mass and pulsation $\omega=m^2$ and whose energy levels are shifted by $-m^2/2$. The spectrum is degenerate in the ``quantum numbers'' $\ell_i$ and labeled by a nonnegative  integer $n$,
\begin{equation}
    \Lambda_{n,\ell} = n m^2 ,
    \label{eq:Gaussianeigenvalues}
\end{equation}
and the eigenfunctions are easily obtained in Cartesian coordinates (in field space) as products of Hermite polynomials. In radial coordinates, they can be written as finite polynomials,
\begin{equation}
    \mathcal{R}_{n,\ell}(x) = e^{-N m^2 x^2/2} r_{n,\ell}(x),
    \label{eq:solgauss}
\end{equation}
where $n-\ell=2k$ is bound to be a nonnegative even integer and
\begin{equation}
     r_{n,\ell}(x) =x^\ell \sum_{q=0}^{\frac{n-\ell}{2}} a_q x^{2q},
    \label{eq:pol}
\end{equation}
is a finite polynomial in $x$ whose coefficients $a_q$ are determined by the recursion relation
\begin{equation}
     (N+2\ell+2q) (q+1)a_{q+1} = -2Nm^2 \qty(k-q) a_{q} .
    \label{eq:rec}
\end{equation}

The latter and, hence, the polynomial in Eq.~\eqref{eq:pol} has a well-defined limit when $N\to\infty$ at fixed $n$ and $\ell$. In this limit, Eq.~\eqref{eq:rec} becomes
\begin{equation}
    (q+1) a_{q+1} =  - 2m^2  \qty(k-q)a_{q} .
    \label{eq:reclargeN}
\end{equation}
which is solved as $a_q=a_0C_{k}^q(-2m^2)^q$, with $C_k^q$ the binomial coefficient, yielding the leading-order radial eigenfunctions, up to a normalization constant,
\begin{equation}
    r_{n,\ell}(x) = a_0x^\ell \qty(1-2 m^2 x^2)^{\frac{n-\ell}2}.
    \label{eq:Gaussianeigenfunctions}
\end{equation}

A few comments are in order here. First, the eigenvalues \eqref{eq:Gaussianeigenvalues} do not scale with $N$. Second, the appropriate radial variable to work with in order to obtain a nontrivial large-$N$ limit is the scaled variable $x$. Finally, taking the limit $N\to\infty$ directly at the level of the eigenvalue equation \eqref{eq:radial} yields the result ${\cal R}=0$ which, although consistent with the naive large-$N$ limit of Eq.~\eqref{eq:solgauss} at fixed $x$, is clearly too harsh. To avoid this caveat, it thus appears important to factor out the exponential factor in \eqref{eq:solgauss}. We now apply these lessons to the case of a more general potential.

\subsection{Interacting case}

Following the previous discussion, we introduce the reduced radial function $\mathcal{R} = e^{-Nv} r$, with $v$ the relevant potential. The eigenvalue equation\eqref{eq:radial} becomes 
\begin{equation}
    -\frac{r''}{2N}  - \qty(\frac{N-1}{2Nx} - v') r' +\frac{\ell(\ell+N-2)}{2N x^2}  r =  \Lambda  r,
    \label{eq:eigeneq}
\end{equation}
which possesses a well-defined large-$N$ limit. Setting $N\to\infty$, we get the following first order equation
\begin{equation}
    \qty(\ln r)' = \frac{ \ell - 2x^2 \Lambda }{x \qty(1 - 2x v')}.
    \label{eq:LOradiallog}
\end{equation}
This can be easily integrated for polynomial potentials in terms of the roots of $(1-2xv')$. We will focus on the case of a quartic potential
\begin{equation}
    v(x) = \frac{m^2}2 x^2 + \frac{\lambda}{4} x^4 ,
    \label{eq:quarticpot}
\end{equation}
 which provides simple analytical formulas.
Using the identity 
\begin{equation}
    1 - 2x v' = \qty(1-2m_+^2 x^2)\qty(1+2m_-^2x^2),
    \label{eq:factor}
\end{equation}
where 
\begin{equation}
    m_\pm^2 = \pm\frac{m^2}2 + \sqrt{\frac{m^4}4 + \frac\lambda2},
    \label{eq:mpm}
\end{equation}
the right-hand side of Eq.~\eqref{eq:LOradiallog} can be decomposed in simple fractions as
\begin{equation}
    \qty(\ln r)'  = \frac{\ell}{x} - \frac{4\alpha_+ m_+^2 x}{1-2m_+^2x^2} +\frac{4\alpha_-  m_-^2 x}{1+2m_-^2 x^2},
    \label{eq:decomp}
\end{equation}
with 
\begin{equation}
    \alpha_\pm = \frac{\pm\Lambda - \ell m_\pm^2}{2\qty(m_+^2 +m_-^2)}.
    \label{}
\end{equation}
Integrating Eq.~\eqref{eq:decomp} is now elementary and yields the leading-order radial function
\begin{equation}
    r(x) = a_0 x^\ell \qty(1-2m_+^2x^2)^{\alpha_+}\qty(1+2m_-^2x^2)^{\alpha_-} .
    \label{}
\end{equation}

The obtained eigenfunctions lead to normalizable PDFs thanks to the exponential factors we have extracted, $P\propto e^{-V}{\cal R}\propto e^{-2V}r$. Requiring the solutions to be regular for all $x$ selects a discrete subset, as expected from the analogous quantum mechanical problem. Using the fact that $m_\pm^2\ge0$, we see that regularity imposes $\alpha_+=k\in\mathds{N}$. In turns, this implies that the eigenvalues are indexed by to nonnegative integers $n$ and $\ell$ such that $n-\ell=2k$ (so that $\ell\le n$) and are given by
\begin{align}
    \Lambda_{n,\ell}= n m_+^2 + (n-\ell) m_-^2.
    \label{eq:LOeigenvalues}
\end{align}
The corresponding eigenfunctions thus read
\begin{equation}
    r_{n,\ell} (x)= a_0 x^\ell \qty(1-2m_+^2x^2)^{\frac{n-\ell}2} \qty(1+2m_-^2x^2)^{-\frac{n}2}.
    \label{eq:LOeigenfunctions}
\end{equation}
Notice that, as expected, the lowest eigenstate of the system has a vanishing eigenvalue, $\Lambda_{0,0}=0$. This is, in fact, guaranteed by the symmetries of the Fokker-Planck operator and simply corresponds to existence of a late-time, equilibrium state of the system, with PDF $P\propto e^{-2V}$. 

Equations \eqref{eq:LOeigenvalues} and \eqref{eq:LOeigenfunctions} completely solve the eigenvalue problem in the large-$N$ limit and provide the leading order of a systematic $1/N$ expansion. As an illustration, we explicitly compute the next-to-leading order corrections in Appendix \ref{sec:NLOcomputation}. We report here the results for the eigenvalues 
\begin{align}
        \Lambda_{n,\ell} &= n m_+^2 + (n-\ell) m_-^2 \nonumber\\
    &+  \frac{\lambda}{2N(m^4+2\lambda)}\left(a_{n,\ell} m_+^2 - b_{n,\ell} m_-^2\right) + \order{\frac1{N^2}},
    \label{eq:NLOeigenvalues}
\end{align}
where $a_{n,\ell} =n(3n-2)-\ell(\ell-2)$ and $b_{n,\ell}=a_{\ell-n,\ell}$.

\section{Discussion}
\label{sec:discussion}

As already mentioned, all physical information in the stochastic approach can be expressed in terms of the eigenvalues and eigenfunctions. The results of the previous section provide analytical expressions, nonperturbative in the coupling constant, which allow one to discuss various physically relevant regimes.
Before doing so, let us quickly mention that the general results of the previous section reproduce the findings of Refs.~\cite{Gautier:2015pca,Moreau:2019jpn}, where the eigenvalues $\Lambda_{1,1}$, $\Lambda_{2,0}$, and $\Lambda_{3,1}$ had been obtained by other means at leading and next-to-leading---for $\Lambda_{1,1}$---orders.
Also worth mentioning is the fact that Eqs.~\eqref{eq:LOeigenvalues} and \eqref{eq:LOeigenfunctions} trivially reduce to the Gaussian results \eqref{eq:Gaussianeigenvalues} and \eqref{eq:Gaussianeigenfunctions} when $\lambda=0$, which corresponds to 
$m_-^2=0$ and $m_+^2=m^2$. 

\begin{figure}[t]
    \centering
    \includegraphics[width=8cm]{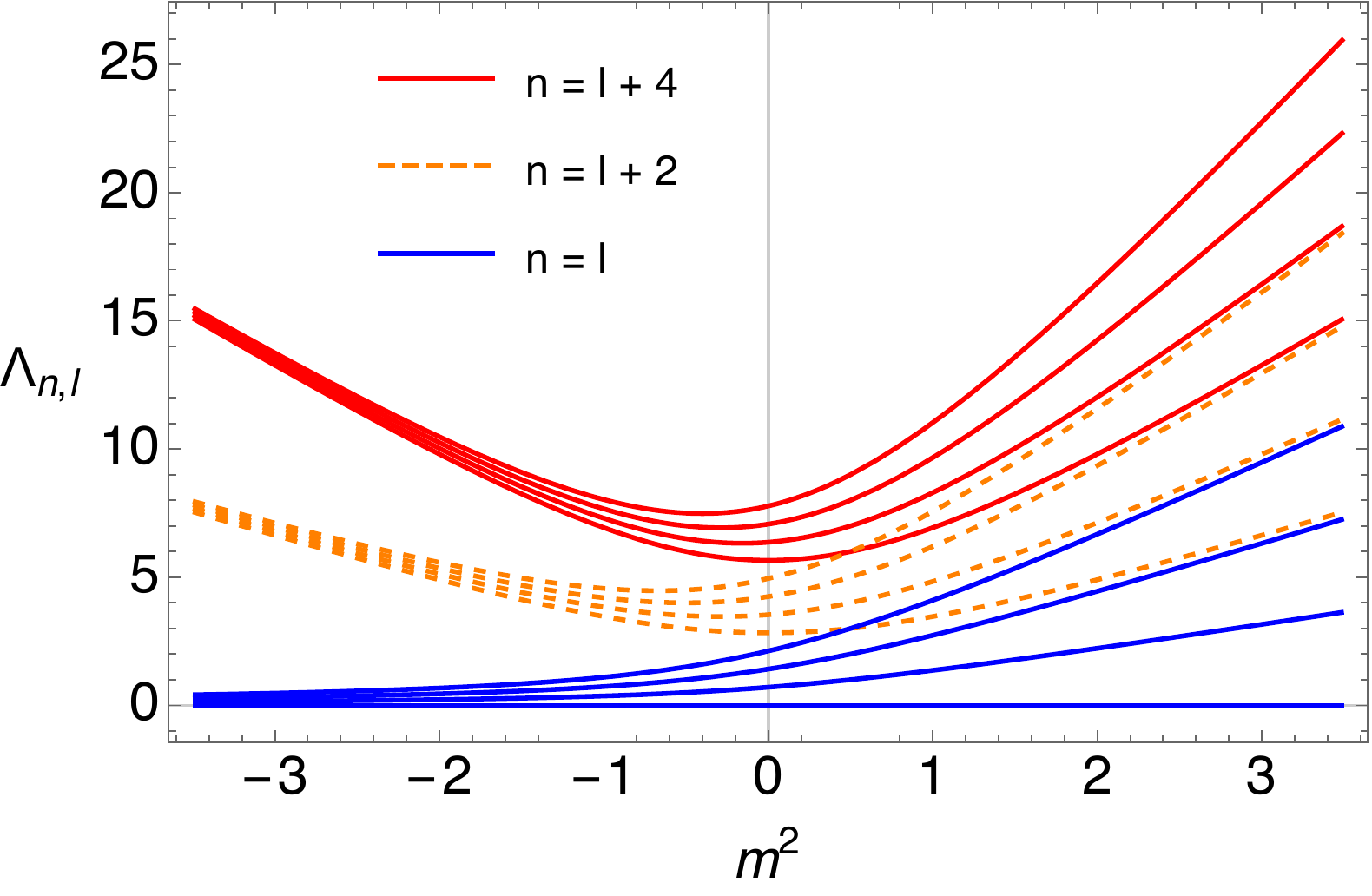}
    \caption{The leading-order  eigenvalues $\Lambda_{n,\ell}$ as functions of $m^2$ at fixed coupling $\lambda=1$.  We show three groups corresponding to (from bottom to top on the axis $m^2=0$) $n=\ell$, $n=\ell+2$, and $n=\ell+4$ with $\ell=0,1,2,3$ for each group.} 
    \label{fig:eigenvalues}
\end{figure}

We show, in Fig.~\ref{fig:eigenvalues}, the evolution of the spectrum of the theory as a function of the parameters of the potential. The Gaussian limit is controlled by the dimensionless coupling $\lambda/m^4$. The lifting of the Gaussian degeneracy, where the eigenvalues $\Lambda_{n,\ell}$ are independent of $\ell$, is given by, in the large-$N$ limit,
\begin{equation}
 \Lambda_{n,\ell}= nm^2+(2n-\ell)\frac{\lambda}{2m^2}+\order{\lambda^2/m^6}.
\end{equation}
For $m^2\to0$, the perturbative treatment is invalid. However, the nonperturbative expressions of the previous section remain valid and give
\begin{equation}\label{eq:massless}
 \Lambda_{n,\ell}\approx(2n-\ell)\sqrt{\frac{\lambda}{2}}.
\end{equation}
The square mass $m_{\rm dyn}^2=\sqrt{\lambda/2}$ is dynamically generated by the self-interactions and is of gravitational origin. It is the so-called dynamical mass, which quantifies the local field fluctuation $\ev{\varphi^2}=1/(2m_{\rm dyn}^2)$. We see that the spectrum \eqref{eq:massless} consists in multiples of this square mass and is thus analog to that of a Gaussian potential with pulsation $m_{\rm dyn}^2$ (although the degeneracies of the eigenvalues are different from the Gaussian case). 

Finally, another interesting regime is that of a double-well potential with $m^2<0$, which is also strongly nonperturbative due to the flat (Goldstone) directions in the potential. We find that the deep-well limit $\lambda/m^4\ll1$ mirrors the near Gaussian case in that $m_+^2\to0$ and $m_-^2\to|m^2|$. In this regime, 
\begin{equation}
 \Lambda_{n,\ell}=(n-\ell)|m^2| +\order{\lambda/m^2}.
\end{equation}
The eigenvalues now only depend on the even integer $n-\ell$ and are thus multiples of $2|m^2|$. This is a simple consequence of the fact that in the deep-well regime, the lowest excitations are those of the approximately Gaussian well of pulsation $2|m^2|$ near the nontrivial minimum. The increased (infinite) degeneracy of each level as compared to the free-field case reflects the presence of flat directions in the potential. In particular, there are infinitely many states with almost zero eigenvalue $\Lambda_{n,n}\approx n\lambda/(2|m^2|)\ll|m^2|\ll1$, which results in large correlation times and lengths for operators in arbitrary nontrivial (vector, tensor, {\it i.e.}, $\ell\neq0$) representation of the symmetry group. The scalar ($\ell=0$) sector is particular in that the only contribution it receives from this light multiplet is the ground state level $\Lambda_{0,0}=0$, which corresponds to the equilibrium PDF and describes the disconnected piece of correlators. The nontrivial time dependence of correlators of scalar operators is thus entirely dictated by the higher levels $\Lambda_{2n,0}\approx 2n|m^2|$, with $n>0$, corresponding to the heavy radial directions in the potential, with square mass $2|m^2|$, and thus by small correlation times and lengths relative to the $\ell\neq0$ sectors. 

\begin{figure}[t]
    \centering
    \includegraphics[width=8.5cm]{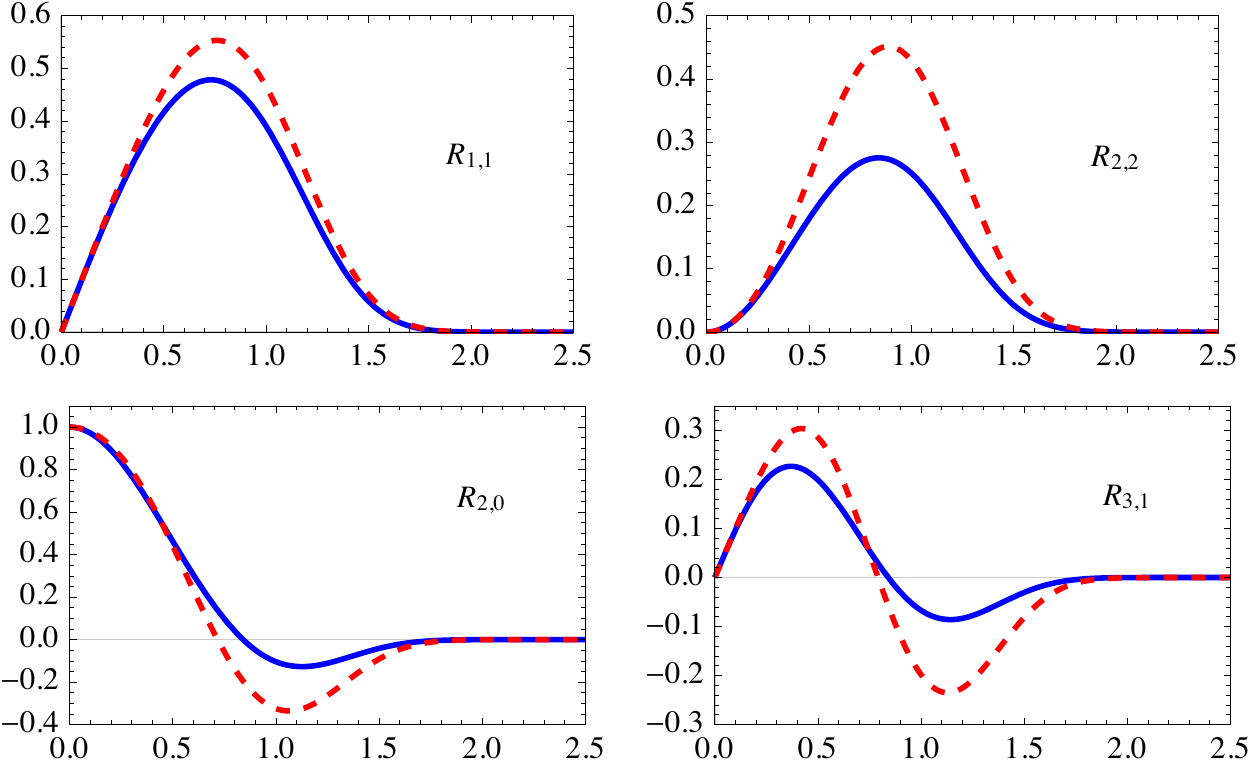}
    \caption{The leading-order (lines) and next-to-leading-order (dashed) eigenfunctions ${\cal R}_{n,\ell}(x)$ for some of the lowest levels in the case $m^2=0$ and $\lambda=1$. We take $N=2$ in this figure to amplify the difference between the two curves in each case. Here, the normalization are chosen such that either the function or its first nonzero derivative at $x=0$ is fixed to $1$. In practice, this means that $a_0=1$ and the function $C(x)$ in Appendix~\ref{sec:NLOcomputation} is chosen such that $C(0)=0$. } 
    \label{fig:eigenfunctions}
\end{figure}

We also show, in Fig.~\ref{fig:eigenfunctions}, the eigenfunctions ${\cal R}_{n,\ell}$ corresponding to some of the lowest eigenstates in the case $m^2=0$. We compare the leading- and next-to-leading-order results (see Appendix \ref{sec:appnum}) with $N=2$ in order to maximise the difference. The explicit expression of the eigenfunctions at next-to-leading order is given in Appendix~\ref{sec:NLOcomputation}.

We end this Section with some comments concerning the applicability of the present results to arbitrary values of $N$. It is often the case that the large-$N$ expansion provides a good---qualitative if not quantitative---guide down to small values of $N$, in particular, in the case $m^2\ge0$. The case $N=1$ has been studied in great detail in the literature \cite{Starobinsky:1994bd,Markkanen:2019kpv,Markkanen:2020bfc} but only very few results exist for $N\ge2$ \cite{Adshead:2020ijf}, to which we compare our findings in the Appendix. In Fig.~\ref{fig:lambda11} , we show the result of numerically computing the lowest eigenvalue $\Lambda_{1,1}$ for various $N$ in the case $m^2=0$ and we compare with the results of the $1/N$ expansion. The leading-order result gives a good qualitative description down to rather low values of $N$ and the first, $1/N$ correction improves the agreement to a quantitative level. We refer the reader to Appendix \ref{sec:appnum} for more details and more comparisons.

\begin{figure}[t]
    \centering
    \includegraphics[width=8cm]{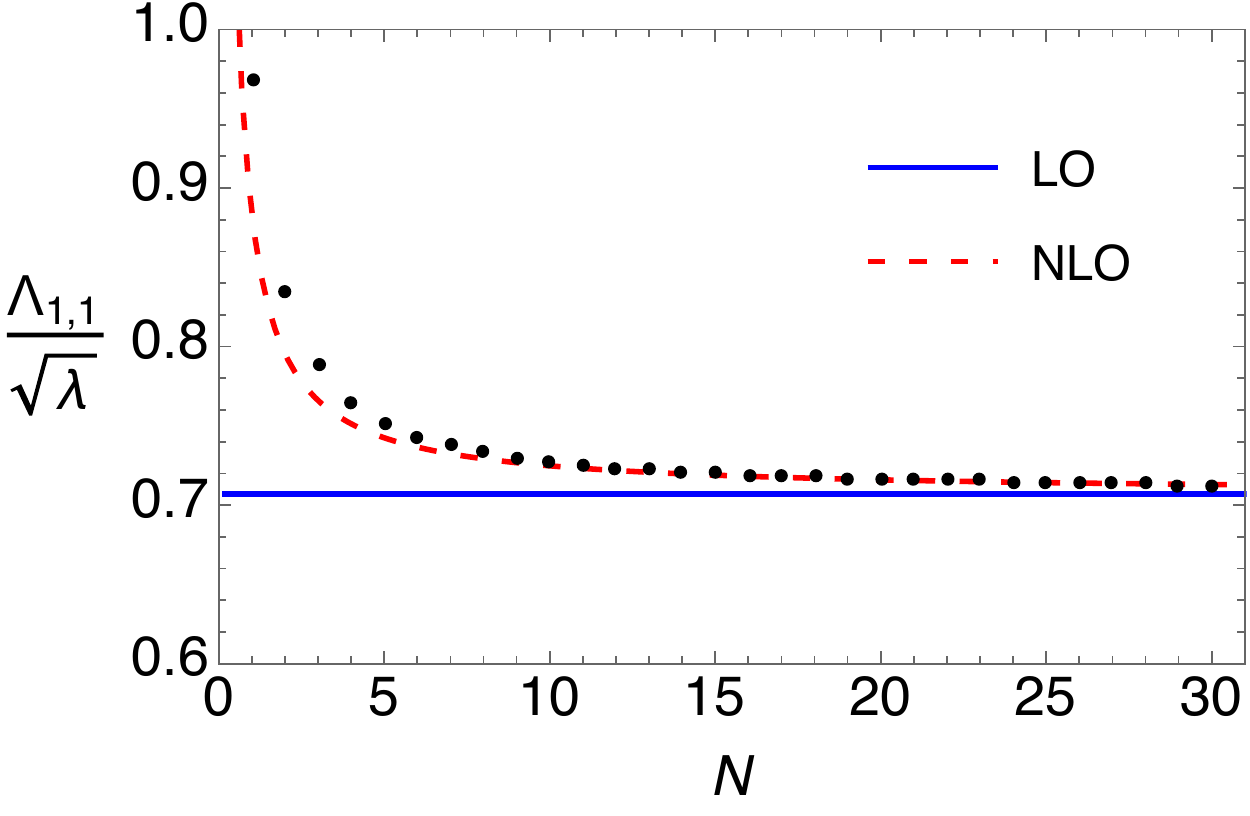}
    \caption{The lowest nonzero eigenvalue $\Lambda_{1,1}$ for the massless case $m^2=0$ as a function of $N$. The dots are the exact values obtained numerically and the curves are the leading-order (LO) and next-to-leading-order (NLO) approximations in the $1/N$ expansion.}
    \label{fig:lambda11}
\end{figure}
Double-well potentials, with $m^2<0$, needs to be discussed separately. First, the case $N=1$ is qualitatively different because the symmetry at work is discrete. In particular, there are no flat directions in the potential and the relevant physics is governed by tunnelling effects, not Goldstone modes. Another aspect which plays an important role for small values of $N$ is the fact that the relevant radial potential for the present eigenvalue problem is not directly $V$, but rather $W$; see Eq.~\eqref{eq:FPp}. Indeed, to reformulate the radial eigenvalue equation \eqref{eq:radial} in terms of a standard one-dimensional problem with an effective potential $W_{\rm eff}$, one eliminates the single derivative term $\propto {\cal R}'$ by means of the redefinition $ \psi(x)=x^{\frac{N-1}2}{\cal R}(x)$. This yields, for arbitrary $N$,
\begin{equation}
   - \frac{1}{2N} \psi'' + W_{\rm eff}\psi=  \Lambda \psi,
    \label{eq:onedschro}
\end{equation}
where
\begin{equation}
    W_{\rm eff}(x) = \frac{\ell(\ell+N-2)}{2Nx^2} + \frac{(N-1)(N-3)}{8Nx^2} + W(x).
    \label{eq:weff}
\end{equation}
Equation \eqref{eq:onedschro} is the standard form of the one-dimensional Schr\"odinger equation with potential $W_{\rm eff}$, except for the factor $N$ in the first term, which can be absorbed in a rescaling of $x$. As Markkanen and Rajantie \cite{Markkanen:2020bfc} pointed out in the case $N=1$ (where $\ell=0,1$ and thus $W_{\rm eff}=W$), in the deep double-well limit, $W$ in fact exhibits a three-well structure as a function of $\varphi$ with, in addition to the symmetric wells at $\varphi\neq0$, a third well around the origin $\varphi=0$. The resulting spectrum is thus, up to exponentially small splittings due to tunnelling effects, a superposition of the Gaussian spectra from the wells at $x=0$ and $x\neq0$, with pulsation $|m^2|$ and $2|m^2|$, respectively, with the bottom of the central well being upshifted by $3|m^2|/2$ relative to that of the external wells.

\begin{figure}[t]
    \centering
    \includegraphics[width=8.5cm]{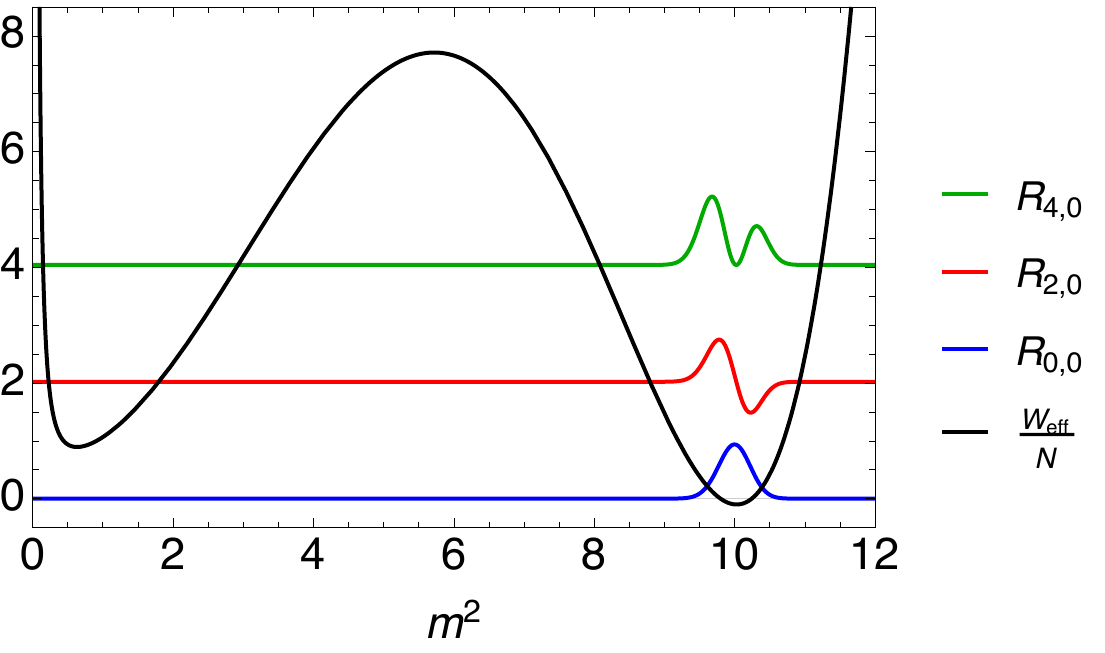}
    \caption{Rescaled effective potential $W_{\rm eff}(x)/N$ with $m^2=-1$, $\lambda=0.01$, and $N=10$, together with the leading-order eigenfunctions $\mathcal{R}_{n,\ell}(x)$ for the first few values of $n$ and $\ell=0$. Although another local minimum appear at low $x$, it is lifted by a factor $N$ and thus gives subleading eigenvalues. The normalization is arbitrary and the eigenfunctions have been upshifted by their respective eigenvalues.}
    \label{fig:Weff}
\end{figure}

The central well remains for arbitrary $N>1$ and the potential \eqref{eq:weff} receives additional centrifugal and geometrical contributions $\propto1/x^2$. Because of the latter and the centrifugal barrier, the minimum of the central well is slightly shifted away from $x=0$ (for $N\ge3$). For increasing $N$, the potential rapidly approaches the asymptotic form
\begin{equation}
    \frac{W_{\rm eff}(x)}{N} =  \frac{1}{8x^2} +   \frac1{2}\qty[ (v')^2- \frac{v'}{x} ]+\order{\frac 1 N}.
    \label{}
\end{equation}
For the quartic potential \eqref{eq:quarticpot}, the position of the central and external wells in the radial direction are given by, respectively, $x_{-}^2=1/(2|m^2|)$ and $x_{+}^2=|m^2|/\lambda$ and the corresponding values of the potential are $W_{\rm eff}(x_-) = 3N|m^2|/4$ and $W_{\rm eff}(x_+)= 0$. We conclude that the excitations of the central well are rapidly lifted relative to that of the external well for increasing $N$ and, hence, decouple in the large-$N$ limit. We show, in Fig.~\ref{fig:Weff}, the effective potential \eqref{eq:weff} together with some of the large-$N$ wavefunctions ${\cal R}_{n,\ell}=e^{-Nv}r_{n,\ell}$, with $r_{n,\ell}$ given by Eq.~\eqref{eq:LOeigenfunctions}.

\section{Conclusions}
\label{sec:concl}

We have proposed a systematic $1/N$-expansion of the stochastic approach for quantum fields in de Sitter spacetime which we have applied to $O(N)$-symmetric models. In its Fokker-Planck formulation, the stochastic approach amounts to solving an equivalent quantum mechanical eigenvalue problem for a single degree of freedom in $N$ dimensions. Various large-$N$ limits of this problem have been considered before in the literature \cite{Dolgov:1979hv,Dutta:1987ce,Chatterjee:1990se} but, to our knowledge, not the one we have proposed here.

We have performed explicit calculations in the case of a quartic potential, for which we have obtained simple analytical expressions of the eigenvalues and eigenfunctions of the Fokker-Planck operator at leading and next-to-leading orders. These reproduce and generalize our previous partial results in Ref.~\cite{Moreau:2019jpn}, where a small subset of eigenvalues could be extracted from the calculation of various correlators in the large-$N$ expansion. The eigenvalues and eigenfunctions obtained here can be used to compute a variety of correlators as well as various time and length scales relevant for both phenomenological and fundamental questions, such as spectral indices, relaxation and decoherence times, etc. \cite{Giraud:2009tn,Gautier:2011fx,Hardwick:2017fjo,Martin:2018lin}. The expressions obtained here are nonperturbative in the relevant coupling constant and are thus useful to analyze the various regimes where a perturbative expansion is unavailable, namely, the cases of light interacting fields, $m^4/H^4\ll\lambda$, and of potentials, with $m^2<0$.

Of course the relevant eigenvalue problem can be exactly solved for a wide variety of potentials by numerical means. First results for multifield systems with continuous symmetries have been presented in Ref.~\cite{Adshead:2020ijf} and in the present work for massless fields with a purely quartic potential. A detailed investigation of more general potentials for various values of $N$---in the spirit of Refs.~\cite{Markkanen:2019kpv,Markkanen:2020bfc} for the case $N=1$---would be of great interest.

\section*{Acknowledgements}
We are thankful to T. Markanen, A. Rajantie, and V. Vennin for useful discussions.  

\appendix

\section{Computation of the eigenvalues and eigenstates at next-to-leading order}
\label{sec:NLOcomputation}

We start by inserting the $1/N$ expansion of the eigenfunctions and eigenvalues, 
\begin{align}
    r &= r_{0} + \frac{r_{1}}{N}+\order{\frac1{N^2}} \\
    \Lambda &= \Lambda_{0} + \frac{\Lambda_{1}}N +\order{\frac1{N^2}},
    \label{}
\end{align}
in Eq.~\eqref{eq:radial}. The leading-order equation is given by Eq.~\eqref{eq:LOradiallog}, and was solved in Sec.~\ref{sec:LOcomputation}. The leading-order eigenfunctions and eigenvalues depend on two quantum numbers $n$ and $\ell$, and are given in Eqs.~\eqref{eq:LOeigenvalues} and \eqref{eq:LOeigenfunctions}. To keep the formulas simple, we will no write explicitly the dependence in the quantum numbers in the following. We define 
\begin{equation}
    g = \frac{\ell - 2x^2 \Lambda_{0}}{x(1 - 2x v')}
    \label{}
\end{equation}

The next-to-leading-order equation reads
\begin{equation}
    \begin{aligned}
     & -\qty(\frac1{2x} - v') r_{1}' + \qty(\frac{\ell}{2x^2} - \Lambda_{0}^{n,\ell}) r_{1} = \\
     & \qquad  \frac{r_{0}''}{2} - \frac{r_{0}'}{2x} - \qty[\frac{\ell(\ell-2)}{2x^2} - \Lambda_1] r_{0} .
    \end{aligned}
    \label{}
\end{equation}
The right-hand side can be written in terms of $g$ using the following relations
\begin{align}
        r_{0}' &= g r_{0},\\
        r_{0}'' &= \qty(g' + g^2) r_{0},
    \label{}
\end{align}
together with the factorization \eqref{eq:factor}, and we end up with
\begin{equation}
    r_{1}' - g r_{1} = hr_{0},
    \label{eq:NLOradial}
\end{equation}
where
\begin{equation}
    h = \frac{x g (1-xg) - x^2 g' + \ell(\ell-2) - 2x^2 \Lambda_{1}}{x(1-2xv')}.
    \label{}
\end{equation}

Using the method of variation of constants, we take the following ansatz, $r_{1}(x) = C(x) r_{0}(x)$, into Eq.~\eqref{eq:NLOradial}, which yields the equation  
\begin{equation}
    C' = h.
    \label{appeq:cprime}
\end{equation}

For the quartic potential \eqref{eq:quarticpot}, the function $h$ is a polynomial fraction which can, again, be decomposed into partial fractions. Introducing the notations 
\begin{equation}
    p_\pm= 1\mp2m_\pm^2x^2,
    \label{}
\end{equation}
with the definition \eqref{eq:mpm}, the functions $r_0$ and $g$ can be expressed as 
\begin{equation}
r_0=a_0x^\ell p_+^{\alpha_+}p_-^{\alpha_+}
\end{equation}
and
\begin{equation}
    g=  \frac{\ell}{x} +\alpha_+  \frac{p_+'}{p_+}+\alpha_- \frac{p_-'}{p_-},
     \label{appeq:g}
\end{equation}
with $\alpha_+=\frac{n-\ell}2$ and $\alpha_-=-\frac n2$. Note also that 
\begin{equation}
 1-2xv'=p_+p_-.
     \label{appeq:pp}
\end{equation}
We obtain, after some calculations,
\begin{equation}
    h = \sum_{k=1}^3 \qty[\alpha_k\frac{ p_+'}{p_+^k} +\beta_k \frac{ p_-'}{p_-^k} ],
    \label{appeq:hdec}
\end{equation}
with the coefficients
\begin{align}
        \alpha_1 &= \frac{\Lambda_{1}}{2M^2} - \frac{m_+^2 m_-^2 (a_{n,\ell} m_+^2 - b_{n,\ell} m_-^2)}{2M^6} ,\\
        \alpha_2 &= (n-\ell)m_+^2\frac{2(n-\ell)m_-^2-(n+\ell-2)m_+^2  }{2M^4},\\
        \alpha_3 &= (n-\ell)(n-\ell-2)\frac{m_+^2}{2M^2} ,
\end{align}
and
\begin{align}
        %\beta_1 &= -\frac{\Lambda_{1}}{2M^2} + \frac{m_+^2 m_-^2 (a_{n,\ell} m_+^2 - b_{n,\ell} m_-^2)}{2M^6} = -\alpha_1, \\
        \beta_1 &= -\alpha_1, \\
        \beta_2 &= n m_-^2\frac{2n m_+^2-(n-2\ell+2)m_-^2 }{2M^4},\\
        \beta_3 &=n(n+2) \frac{m_-^2}{2M^2} ,
    \label{}
\end{align}
where we defined $M^2=m_+^2+m_-^2$ and
\begin{align}
        a_{n,\ell} &=n(3n-2)-\ell(\ell-2)\\
        b_{n,\ell} &=(n-\ell)(3n-3\ell+2)-\ell(\ell-2).
    \label{}
\end{align}
Note that $b_{n,\ell} =a_{\ell-n,\ell}$. We verify explicitly the $+\leftrightarrow-$ symmetry, obvious from Eqs.~\eqref{appeq:g} and \eqref{appeq:pp}. In particular, we check that the coefficients $\alpha_k\leftrightarrow\beta_k$ under the exchange $m_+^2\leftrightarrow -m_-^2$ and $n-\ell\leftrightarrow-n$.

With the decomposition \eqref{appeq:hdec}, Eq.~\eqref{appeq:cprime} is readily integrated as 
\begin{align}
    C &=  \alpha_1 \log \frac{p_+}{p_-} - \frac{\alpha_2}{p_+} - \frac{\alpha_3}{2p_+^2} \nonumber\\
    & \quad   - \frac{\beta_2}{p_-} - \frac{\beta_3}{2p_-^2}+a_1,
    \label{appeq:Cfunc}
\end{align}
with $a_1$ a free integration constant to be fixed, {\it e.g.}, by a normalization condition at next-to-leading order. As before, at leading order, possible singularities are related to the zero of the polynomial $p_+$. Remembering that the solution we seek is $r_1=Cr_0$, we see that the last two terms in the first line of Eq.~\eqref{appeq:Cfunc} contribute as  $\alpha_2p_+^{\alpha_+-1}$ and $\alpha_3p_+^{\alpha_+-2}$ and are thus potentially singular for $n-\ell=0$ and $n-\ell=0,2$, respectively. This singularities are, in fact, absent thanks to the fact that the coefficients $\alpha_2$ and $\alpha_3$ vanish for these values of $n-\ell$. The only possible singular behavior comes from the term $\ln p_+$ and regularity thus imposes $\alpha_1=0$. This  fixes $\Lambda_{1}$ as 
\begin{equation}
    \Lambda_{1} = \frac{m_+^2 m_-^2}{M^4} \qty(a_{n,\ell} m_+^2 - b_{n,\ell} m_-^2) .
    \label{}
\end{equation}
The expression \eqref{eq:NLOeigenvalues} is obtained using 
the identities  $M^2=\sqrt{m^4+2\lambda}$ and $m_+^2m_-^2=\lambda/2$. Finally, the corresponding eigenfunction reads
\begin{equation}
        r_{n,\ell}(x)=  \qty[1+\frac{C_{n,\ell}(x)}{N}+\order{\frac1{N^2}}]r^{n,\ell}_0(x).
    \label{}
\end{equation}

The above expressions are valid for all values of the parameters $m^2$ and $\lambda$. To end this Section, we present the explicit formulas for the case $m^2=0$,  where $m_+^2=m_-^2=\sqrt{\lambda/2}$. We have
\begin{align}
    \frac{\Lambda_{n,\ell}}{\sqrt{\lambda}} =  \frac{2n-\ell}{\sqrt{2}}\left[1+\frac{3\ell-2}{4N}+\order{\frac1{N^2}}\right]
    \label{appeq:NLOmassless}
\end{align}
and the various coefficients in the function $C(x)$ read
\begin{align}
        \alpha_2 &= \frac{(n-\ell)(n-3\ell+2)}{8}\\
        \alpha_3 &= \frac{(n-\ell)(n-\ell-2)}{4} \\
        %\beta_1 &= \frac{(2n-\ell)(3\ell-2)}{8} \\
        \beta_2 &= \frac{n(n+2\ell-2) }{8}\\
        \beta_3 &= \frac{n(n+2)}{4} .
    \label{}
\end{align}
As an illustration, the corresponding eigenfunctions are plotted against the leading-order ones in Fig.~\ref{fig:eigenfunctions} for $N=2$. In practice, we observe that the next-to-leading-order eigenfunctions provide a pretty good approximation of the numerical results down to $N=2$ for the eigenstates we have computed numerically here, namely, ${\cal R}_{1,1}$ and ${\cal R}_{2,0}$.

\section{Comparison with numerical results}\label{sec:appnum}

In order to test the validity and convergence of the $1/N$ expansion in a simple---but nontrivial---case, we solve numerically the eigenvalue equation \eqref{eq:eigeneq} for a purely quartic potential \eqref{eq:quarticpot}, with $m^2=0$. We first compute the lowest nonzero eigenvalue $\Lambda_{1,1}$
as a function of $N$ and compare with the leading and next-to-leading-order predictions \eqref{appeq:NLOmassless}. This is presented in Fig.~\ref{fig:lambda11}.
The first observation is that the leading-order result gives a reasonable estimate of the exact result down to rather low values of $N$. Furthermore, the next-to-leading-order approximation neatly improves the matters and gives a fairly accurate description of the exact results down to $N=1$, where the relative error is $8\%$. From Eq.~\eqref{appeq:NLOmassless}, we also observe that the vector ($\ell=1$) sector is the one with the smallest $1/N$ correction. 

\begin{figure}[t!]
    \centering
    \includegraphics[width=8cm]{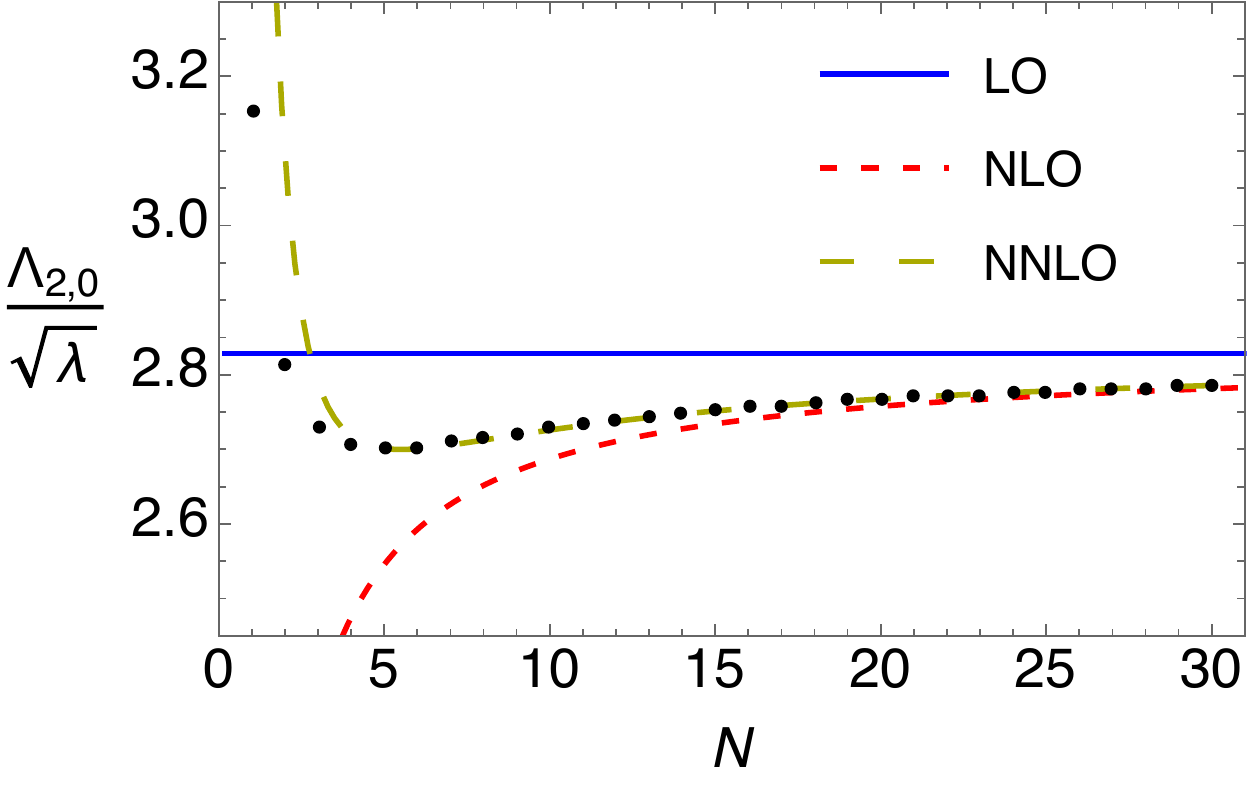}
    \caption{Same as Fig.~\ref{fig:lambda11} for the lowest nonzero eigenvalue in the scalar sector, $\Lambda_{2,0}$. Here, we also show the next-to-next-to-leading-order result (NNLO).}
    \label{fig:lambda20}
\end{figure}

To test further the present expansion scheme, we do the same analysis for the lowest nonzero eigenvalue in the scalar ($\ell=0$) sector, namely $\Lambda_{2,0}$. This is presented in Fig.~\ref{fig:lambda20}. The leading-order result gives, again, a good estimate of the exact result down to low values of $N$.  As before, the next-to-leading-order approximation quantitatively improves the description, however, for not too small values of $N$, for which it becomes worse. The relative error reaches $8\%$ for $N=4$ and increases up to $25\%$ for $N=1$. We have computed the $1/N^2$ correction in that case, which reads
\begin{equation}
 \frac{\Lambda_{2k,0}}{\sqrt{\lambda}}=\frac{4k}{\sqrt{2}}\left[1-\frac{1}{2N}+\frac{1+10k^2}{8N^2}+\order{\frac1{N^3}}\right]
\end{equation}
and is also shown in Fig.~\ref{fig:lambda20}. We see that it greatly improves the description at small $N$, with a relative error of $10\%$ for $N=1$.

We end this Section by comparing, when possible, our numerical results to existing ones in the literature. The case $N=1$ has been studied in great detail in Refs.~\cite{Starobinsky:1994bd,Markkanen:2019kpv,Markkanen:2020bfc} and, recently, Adshead {\it et al} have presented first results for continuous symmetries, with $N=2,3$ in the case of a purely quartic potential \cite{Adshead:2020ijf}. The quartic coupling $\tilde\lambda$ in that Reference is related to ours as $\tilde\lambda =d^2\Omega_{D+1}\lambda/(2N)$ and the authors use the quantum numbers $k=(n-\ell)/2$ and $\ell$ to label the eigenstates. Their definition of the eigenvalues $\tilde\Lambda_{k,\ell}$ is related to ours as
\begin{equation}
 \tilde\Lambda_{k,\ell}=\frac{\Lambda_{2k+\ell,\ell}}{\sqrt\lambda}\sqrt{\frac{N}{12\pi^2}},
\end{equation}
with $d=3$ and $\Omega_5=8\pi^2/3$. This holds for $N\ge2$. In the case $N=1$, there is no angular momentum in field space and only $\ell=0,1$ are permitted. One has
\begin{align}
 \tilde\Lambda_{2k}&=\frac{\Lambda_{2k,0}}{\sqrt\lambda}\sqrt{\frac{1}{12\pi^2}}\\
 \tilde\Lambda_{2k+1}&=\frac{\Lambda_{2k+1,1}}{\sqrt\lambda}\sqrt{\frac{1}{12\pi^2}}.
\end{align}
For $N=1$, we find $\Lambda_{1,1}/\sqrt\lambda=0.9693$ and $\Lambda_{2,0}/\sqrt\lambda=3.1532$. This translates into $\tilde \Lambda_{1}=0.0891$ and $\tilde \Lambda_{2}=0.2897$, which agrees with the known results \cite{Starobinsky:1994bd,Markkanen:2019kpv,Markkanen:2020bfc,Adshead:2020ijf}. For cases with continuous symmetries, we find $\Lambda_{2,0}/\sqrt\lambda=2.8133$ for $N=2$ and $\Lambda_{2,0}/\sqrt\lambda=2.7296$ for $N=3$, which give $\tilde \Lambda^{N=2}_{1,0}=0.3656$ and $\tilde \Lambda^{N=3}_{1,0}=0.4344$, respectively, in agreement with the results of Ref.~\cite{Adshead:2020ijf}.

\end{document}